\documentclass[prd,aps,preprintnumbers,floatfix,showpacs,preprint,tightenlines]{revtex4}
\usepackage{epsfig}
\usepackage{graphics}
\usepackage{latexsym}
\usepackage{amsmath}
\usepackage{amssymb}
\usepackage{rotating}

\begin{document}

~~~~~~~~~~~~~~~~~~~~~~~~~~~~~~~~~~~~~~~~~~~~~~~~~~~~~~~~~~~~~~~~~~~~~~~~~~~~~~~~~~~~~~~~~ADP-13-34/T854

\title{Strange magnetic form factor of nucleon in heavy baryon chiral effective approach at next to leading order}

\author{P. Wang$^{ab}$}
\author{D. B. Leinweber$^c$}
\author{A. W. Thomas$^{cd}$}

\affiliation{$^a$Institute of High Energy Physics, CAS, P. O. Box
918(4), Beijing 100049, China}

\affiliation{$^b$Theoretical Physics Center for Science Facilities,
CAS, Beijing 100049, China}

\affiliation{ $^c$Special Research Center for the Subatomic Structure
  of Matter (CSSM), School of Chemistry \& Physics, University of
  Adelaide, SA 5005, Australia}

\affiliation{ $^d$ ARC Centre of Excellence in Particle Physics
at the Terascale,
School of Chemistry \& Physics, University of Adelaide, SA 5005, Australia}

\begin{abstract}
The strange magnetic form factor of the nucleon is studied in the heavy baryon chiral effective approach at next to leading order.
The one loop contributions from kaon and intermediate octet and decuplet hyperons are included,
using finite-range-regularization to deal with the ultra-violet divergences.
Drawing on an established connection between quenched and full QCD, this model makes it possible to predict the strange magnetic form factor under the
hypothesis that for a dipole regulator mass $\Lambda$ around 0.8 GeV, strangeness in the core is
negligible. The strange magnetic form factor is found to be small and negative over a range of momentum transfer,
while the strange magnetic moment is consistent with the best lattice QCD
determinations.
%while the strange magnetic moment is $G_M^s(0)=-0.058^{+0.034}_{-0.053}$.

\end{abstract}


\maketitle

\section{Introduction}
While our understanding of Quantum Chromodynamics (QCD) in
the perturbative sector is excellent, in the
non-perturbative sector we have much to learn. Certainly lattice QCD
has achieved a great deal of success and the masses of the ground state hadrons
are under control~\cite{Durr:2008zz,Young:2009zb}.
The detailed study of excited states is
beginning in earnest~\cite{Mahbub:2012ri,Edwards:2011jj}
and there has been some success in the calculation
of elastic form factors~\cite{Schroers:2009zc,Collins:2011mk,Leinweber:1990dv,Leinweber:1992hy,Boinepalli:2006xd,Boinepalli:2009sq} and
even transition form factors~\cite{Lin:2008qv,Leinweber:1992pv}.
However, with very few exceptions, the form factor studies,
which complement the recent experimental progress at facilities such as Jefferson Lab,
deal with so-called ``connected contributions" in which the external current
acts on a quark line running directly from the hadronic source to sink. As discussed below,
only a very few studies have directly addressed the ``disconnected contributions''.

Perhaps the most famous example of a disconnected contribution is the strange quark
contribution to the nucleon elastic form factors~\cite{Leinweber:1999nf,Thomas:2012tg}.
Its fundamental importance is
associated with the fact that it is directly analogous to the vacuum polarization
contribution to the Lamb shift, the correct calculation of which confirmed the
validity of Quantum Electrodynamics. If QCD is to be confirmed unambiguously as the
complete theory of QCD and our capacity to use it judged reliable, it is crucial
that it successfully reproduce the measured strange quark form factors.

The discovery of the EMC spin crisis~\cite{Ashman:1987hv}
in the late 80's stimulated enormous
interest in the strange quark contribution to the nucleon spin, $\Delta s$
(essentially
the matrix element of $\bar{s} \gamma^\mu \gamma_5 s$ in the proton), because of
speculation that the explanation might lie in an unexpectedly large strange
quark spin contribution. Recent lattice QCD studies suggest that $\Delta s$ is
actually rather small~\cite{QCDSF:2011aa},
consistent with a recent model study suggesting the SU(3)
is likely broken at the 20\% level for the axial charge $g_A^8$ of
the nucleon~\cite{Bass:2009ed}.
Instead it appears that the resolution of the spin crisis lies in the exchange of
quark spin for orbital angular momentum of the quarks
(and anti-quarks)~\cite{Myhrer:2007cf,Thomas:2008bd,Thomas:2008ga}. Of
course, this leaves as a facinating challenge the detailed partition of the
nucleon spin amongst the quarks and gluons.

Although the spin crisis has faded as a motivation for detecting the role
strange quarks play in the nucleon, its fundamental role as a totally disconnected
contribution means that it provides a vital role to test our understanding
of QCD~\cite{Leinweber:1999nf}.
An extremely impressive program of measurements of parity violating
electron scattering (PVES) at MIT-Bates~\cite{sample1,sample2},
JLab~\cite{happex1,G01,G02,happex2,happex3} and Mainz~\cite{a41,a42,a43} has provided the
essential third constraint so that, in combination with the electric or magnetic
form factors of the proton and neutron and the assumption of charge symmetry,
a careful global analysis~\cite{Ross1} has allowed a determination of the strange quark
magnetic moment and charge radius. For details of the experimental programs and some related
theoretical work we refer to Refs.~\cite{Armstrong}.
It is a remarkable confirmation of our growing capacity to compute with
non-perturbative QCD that the results
agree well with the most recent determinations from
lattice QCD~\cite{Leinweber:2004tc,Leinweber:2006ug,Doi:2009sq}.

In the future, we can expect that both lattice QCD and experiment will achieve greater
precision at non-zero momentum transfer and so map out the $Q^2$-dependence of
the strange form factors. At the same time, in order to develop a physical
understanding of these results, it will be helpful to have calculations
within a variety of models, informed as far as possible by the lattice QCD results.
In the past, a variety of theoretical models have been applied to the calculation of the
strange nucleon form factors \cite{Zou,Jaffe,Hammer,Cohen,Park,Weigel,Goeke,Hannelius1,Lyubovitskij,Bijker,Kiswandhi,Leinweber:1999nf}.
In 2003, Lewis $et$ $al$.~\cite{Lewis:2002ix} used lowest order, quenched chiral perturbation theory, together with lattice QCD simulations to
calculate the strange form factors. The magnetic form factor which they obtained at $Q^2=0.1$ GeV$^2$ was $+0.05\pm
0.06$. Recently, by combining the constraints of charge symmetry \cite{Miller:1997ya,Miller:1990iz} with new chiral extrapolation techniques and
low mass, quenched lattice QCD simulations of the individual quark contributions to the magnetic moment and form factor of the nucleon
precise, nonzero values of strange magnetic moment and form factor were obtained \cite{Leinweber:2004tc,Wang1}.

Chiral perturbation theory ($\chi$PT) is also a powerful tool with which
to study hadron properties
at low energy. There has been some work on strange form factors with heavy baryon chiral perturbation theory~\cite{Hemmert1,Hemmert2}.
However, there is an unknown low energy constant appearing in the chiral Lagrangian which
has limited the capacity to calculate the strange magnetic form factor.
In other words, the quantity one wishes to predict --
the strangeness vector current matrix element -- is the same
quantity one needs to know in order to make a prediction \cite{Musolf,Kubis}.
While this is the case in conventional $\chi$PT,
experience with finite-range-regularization (FRR),
has shown that by varying the regulator parameter,
one can shift strength from the loop contributions into the core.
This suggests that within FRR $\chi$-EFT one might identify the core contribution with the tree level contribution and make the
approximation that, for $\Lambda$ around 0.8 GeV, the strangeness content of the core is negligible.
In this way, full QCD results have been obtained rather successfully from quenched lattice data \cite{Leinweber:2004tc,Leinweber:2006ug,Wang1,Wang2,Wang3}.
We should emphasize that unquenching only works for $\Lambda$ around 0.8 GeV. Only then does one define a core contribution that is approximately
invariant between quenched and full QCD.

In earlier work, the strange magnetic moment and form factor have been studied at leading
order~\cite{Leinweber:2004tc,Wang1}.
Here we extend the study to next to leading order in order to investigate the role of high-order terms.
It is also important to see to what extent FRR suppresses the high-order contribution to the strange magnetic form factor.
For the strange magnetic form factor with regulator $\Lambda=0.8$ GeV,
the low energy constant related to the 3-quark core contribution is approximately zero.
Therefore, we can make a prediction for the strange magnetic form factor with FRR without any parameter.
The paper is organized as follows. In section II, we briefly introduce the heavy baryon chiral Lagrangian which will
be used in the calculation. The strange magnetic form factor is presented in section III. Numerical results are
shown in section IV. Finally, section V presents a summary of our results.

\section{Chiral Lagrangian}
There are many papers which deal with heavy baryon chiral perturbation
theory -- for details see, for example, Refs.
\cite{Jenkins2,Labrenz,Tiburzi:2004mv}. For completeness, we briefly
introduce the formalism in this section. In heavy baryon chiral
perturbation theory, the lowest chiral Lagrangian for the baryon-meson
interaction which will be used in the calculation of the nucleon
magnetic moments, including the octet and decuplet baryons, is
expressed as
\begin{eqnarray}\label{lol}
{\cal L}_v &=&i{\rm Tr}\bar{B}_v(v\cdot {\cal D})
B_v+2D{\rm Tr}\bar{B}_v S_v^\mu\{A_\mu,B_v\}
+2F{\rm Tr}\bar{B}_v S_v^\mu[A_\mu,B_v]
\nonumber \\
&& -i\bar{T}_v^\mu(v\cdot {\cal D})T_{v\mu}
+{\cal C}(\bar{T}_v^\mu A_\mu B_v+\bar{B}_v A_\mu T_v^\mu),
\end{eqnarray}
where $S_\mu$ is the covariant spin-operator defined as
\begin{equation}
S_v^\mu=\frac i2\gamma^5\sigma^{\mu\nu}v_\nu.
\end{equation}
Here, $v^\nu$ is the nucleon four velocity (in the rest frame, we have
$v^\nu=(1,0)$).
D, F and $\cal C$ are the usual coupling constants.
The chiral covariant derivative, $D_\mu$, is written as $D_\mu
B_v=\partial_\mu B_v+[V_\mu,B_v]$. The pseudoscalar meson octet
couples to the baryon field through the vector and axial vector
combinations
\begin{equation}
V_\mu=\frac12(\zeta\partial_\mu\zeta^\dag+\zeta^\dag\partial_\mu\zeta),~~~~~~
A_\mu=\frac12(\zeta\partial_\mu\zeta^\dag-\zeta^\dag\partial_\mu\zeta),
\end{equation}
where
\begin{equation}
\zeta=e^{i\phi/f}, ~~~~~~
f=93~{\rm MeV}.
\end{equation}
The matrix of pseudoscalar fields $\phi$ is expressed as
\begin{eqnarray}
\phi=\frac1{\sqrt{2}}\left(
\begin{array}{lcr}
\frac1{\sqrt{2}}\pi^0+\frac1{\sqrt{6}}\eta & \pi^+ & K^+ \\
\pi^- & -\frac1{\sqrt{2}}\pi^0+\frac1{\sqrt{6}}\eta & K^0 \\
K^- & \bar{K}^0 & -\frac2{\sqrt{6}}\eta
\end{array}
\right).
\end{eqnarray}
$B_v$ and $T^\mu_v$ are the new, velocity dependent fields
which are related to the original baryon octet and decuplet fields
$B$ and $T^\mu$ by
\begin{equation}
B_v(x)=e^{im_N \not v v_\mu x^\mu} B(x),
\end{equation}
\begin{equation}
T^\mu_v(x)=e^{im_N \not v v_\mu x^\mu} T^\mu(x).
\end{equation}
In the chiral $SU(3)$ limit, the octet baryons will have the same
mass, $m_B$. In our calculation, we use the physical masses for
the members of the baryon octet and decuplet. The explicit form of the baryon octet
is written as
\begin{eqnarray}
B=\left(
\begin{array}{lcr}
\frac1{\sqrt{2}}\Sigma^0+\frac1{\sqrt{6}}\Lambda &
\Sigma^+ & p \\
\Sigma^- & -\frac1{\sqrt{2}}\Sigma^0+\frac1{\sqrt{6}}\Lambda & n \\
\Xi^- & \Xi^0 & -\frac2{\sqrt{6}}\Lambda
\end{array}
\right).
\end{eqnarray}
{}For the baryon decuplets, there are three indices, defined as
\begin{eqnarray}
T_{111}=\Delta^{++}, ~~ T_{112}=\frac1{\sqrt{3}}\Delta^+, ~~
T_{122}=\frac1{\sqrt{3}}\Delta^0, \\ \nonumber
T_{222}=\Delta^-, ~~ T_{113}=\frac1{\sqrt{3}}\Sigma^{\ast,+}, ~~
T_{123}=\frac1{\sqrt{6}}\Sigma^{\ast,0}, \\ \nonumber
T_{223}=\frac1{\sqrt{3}}\Sigma^{\ast,-}, ~~
T_{133}=\frac1{\sqrt{3}}\Xi^{\ast,0}, ~~ T_{233}=\frac1{\sqrt{3}}\Xi^{\ast,-},
~~ T_{333}=\Omega^{-}.
\end{eqnarray}

From the Lagrangian of electromagnetic moments, one can get the strange quark contribution.
The octet, decuplet and octet-decuplet transition magnetic moment
operators are needed in the one loop calculation of the nucleon strange
form factor. The baryon octet magnetic Lagrangian is written as:
\begin{equation}\label{lomag}
{\cal L}=\frac{e}{4m_N}\left(\mu_D{\rm Tr}\bar{B}_v \sigma^{\mu\nu}
\left\{F^+_{\mu\nu},B_v\right\}+\mu_F{\rm Tr}\bar{B}_v \sigma^{\mu\nu}
\left[F^+_{\mu\nu},B_v \right]\right),
\end{equation}
where
\begin{equation}
F^+_{\mu\nu}=\frac12\left(\zeta^\dag F_{\mu\nu}Q\zeta+\zeta
F_{\mu\nu}Q\zeta^\dag\right).
\end{equation}
$Q$ is the charge matrix $Q=$diag$\{2/3,-1/3,-1/3\}$. At the lowest
order, the Lagrangian will generate the following nucleon magnetic
moments:
\begin{equation}\label{treemag}
\mu_p=\frac13\mu_D+\mu_F,~~~~~~ \mu_n=-\frac23\mu_D.
\end{equation}

The decuplet magnetic moment operator is expressed as
\begin{equation}
{\cal L}=-i\frac{e}{m_N}\mu_C q_{ijk}\bar{T}^\mu_{v,ikl}T^\nu_{v,jkl}
F_{\mu\nu},
\end{equation}
where $q_{ijk}$ and $q_{ijk}\mu_C$ are the charge and magnetic
moment of the decuplet baryon $T_{ijk}$.
The transition magnetic operator is
\begin{equation}
{\cal L}=i\frac{e}{2m_N}\mu_T F_{\mu\nu}\left(\epsilon_{ijk}Q^i_l \bar{B}^j_{vm} S^\mu_v
T^{\nu,klm}_v+\epsilon^{ijk}Q^l_i \bar{T}^\mu_{v,klm} S^\nu_v B^m_{vj}\right).
\end{equation}
The electromagnetic moments of baryon octet and decuplet as well as the transition moments
can be written in terms of $\mu_u$, $\mu_d$ and
$\mu_s$ instead of the $\mu_D$, $\mu_F$, $\mu_C$ and $\mu_T$ \cite{Ha1}. For the particular
choice, $\mu_s=\mu_d=-\frac12 \mu_u$, one finds the following
relationship:
\begin{equation}
\mu_D=\frac32 \mu_u, ~~~ \mu_F=\frac23 \mu_D, ~~~ \mu_C=\mu_D, ~~~ \mu_T=-4\mu_D.
\end{equation}
Therefore, by comparing with the quark model, one can get the strange quark contribution
to the magnetic moment of baryons at tree level as \cite{Ha1}
\begin{equation}
\mu_p^s=\mu_n^s=0, ~~~ \mu_{\Sigma^+}^s = \mu_{\Sigma^-}^s=\mu_{\Sigma^0}^s=\mu_s=-\frac13\mu_D, ~~~
\mu_{\Lambda}^s = -3\mu_s = \mu_D.
\end{equation}

Similarly, the strange quark contribution to the decuplet and transition
magnetic moments can be written as \cite{Ha2}
\begin{equation}
\mu_{\Sigma^{*,+}}^s=\mu_{\Sigma^{*,0}}^s=\mu_{\Sigma^{*,-}}^s= - 3\mu_s = \mu_D,
\end{equation}
\begin{equation}
\mu_{\Sigma^{*,+}\Sigma^+}^s=-\mu_{\Sigma^{*,0}\Sigma^0}^s=-\mu_{\Sigma^{*,-}\Sigma^-}^s= -2\sqrt{2}\mu_s = \frac{2\sqrt{2}}{3}\mu_D,
\end{equation}
In the above Eqs.~(16) through (18), following the usual convention, the charge of the strange quark is taken to be 1.
Therefore the standard $\mu_s$ has been multiplied by $-3$.

In the heavy baryon formalism, the propagators of the octet or
decuplet baryon, $j$, are expressed as
\begin{equation}
\frac i {v\cdot k-\delta^{jN}+i\varepsilon} ~~{\rm and}~~ \frac {iP^{\mu\nu}}
{v\cdot k-\delta^{jN}+i\varepsilon},
\end{equation}
where $P^{\mu\nu}$ is $v^\mu v^\nu-g^{\mu\nu}-(4/3)S_v^\mu S_v^\nu$.
$\delta^{ab}=m_b-m_a$ is the mass difference of between the two
baryons. The propagator of the $K$ meson is the usual
free propagator, i.e.:
\begin{equation}
\frac i {k^2-M_K^2+i\varepsilon}.
\end{equation}

\begin{center}
\begin{figure}[tbp]
\includegraphics[scale=0.45]{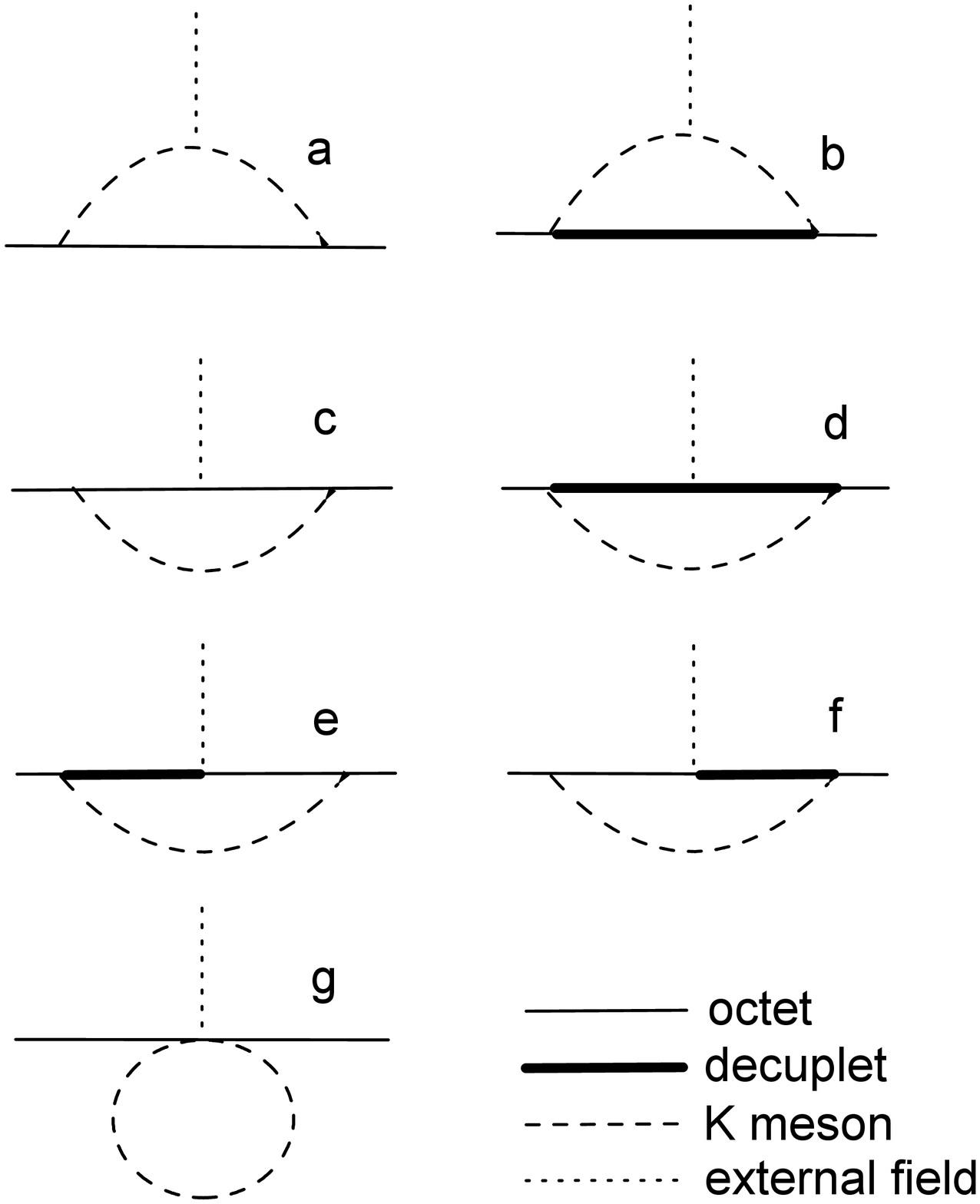}
\caption{The one loop Feynman diagrams for the nucleon strange magnetic form factor.
The solid, thick solid, dash and dotted lines are for the octet baryons,
decuplet baryons, K mesons, and external fields, respectively.}
\end{figure}
\end{center}

\section{Strange Magnetic Form Factor}

The one loop Feynman diagrams which contribute to the strange
magnetic form factor of the nucleon are plotted in Fig.~1. The contribution from Fig.~1a is expressed as
\begin{equation}\label{p1a}
G_M^{s(1a)}=-\frac{m_N(D+3F)^2I_{1K}^{N\Lambda}+9m_N(D-F)^2I_{1K}^{N\Sigma}}{{48\pi^3f_\pi^2}},
\end{equation}
The integration $I_{1K}^{\alpha\beta}$ is expressed as
\begin{equation}
I_{1K}^{\alpha\beta}=\int d\overrightarrow{k}\frac{k_y^2
u(\overrightarrow{k}+\overrightarrow{q}/2)
u(\overrightarrow{k}-\overrightarrow{q}/2)(\omega_K(\overrightarrow{k}+\overrightarrow{q}/2)
+\omega_K(\overrightarrow{k}-\overrightarrow{q}/2)+\delta^{\alpha\beta})}
{A_K^{\alpha\beta}},
\end{equation}
where
\begin{eqnarray}
A_K^{\alpha\beta}&=&\omega_K(\overrightarrow{k}+\overrightarrow{q}/2)
\omega_K(\overrightarrow{k}-\overrightarrow{q}/2)
(\omega_K(\overrightarrow{k}+\overrightarrow{q}/2)+\delta^{\alpha\beta})
\nonumber \\
&&(\omega_K(\overrightarrow{k}-\overrightarrow{q}/2)+\delta^{\alpha\beta})
(\omega_K(\overrightarrow{k}+\overrightarrow{q}/2)+\omega_K(\overrightarrow{k}-\overrightarrow{q}/2)).
\end{eqnarray}
$\omega_K(\overrightarrow{k})=\sqrt{m_K^2+\overrightarrow{k}^2}$ is
the energy of kaon. In our calculation we use the finite-range-regularization and $u(\overrightarrow{k})$ is the ultra-violet
regulator. The first term in Eq.~(\ref{p1a}) comes from the charged
$K$ meson cloud with intermediate $\Lambda$ hyperon. The second term includes both charged and neutral $K$ meson cloud
contributions with an intermediate $\Sigma$ hyperon. Fig.~1b is the same as Fig.~1a but the intermediate
states are decuplet baryons.  Its contribution to the strange magnetic form
factor is expressed as
\begin{equation}
G_M^{s(1b)}=\frac{m_N{\cal
C}^2}{48\pi^3f_\pi^2}I_{1K}^{N\Sigma^\ast}.
\end{equation}
The contribution to the form factor from Fig.~1c is expressed as
\begin{equation} \nonumber
G_M^{s(1c)}=-\frac{1}{192\pi^3f_\pi^2}\left[3(D-F)^2\mu_s
I_{2K}^{N\Sigma}-(D+3F)^2\mu_sI_{2K}^{N\Lambda}\right],
\end{equation}
where
\begin{equation}
I_{2K}^{\alpha\beta}=\int d\overrightarrow{k}\frac{k^2 u(\overrightarrow{k})^2}
{\omega_K(\overrightarrow{k})(\omega_K(\overrightarrow{k})+\delta^{\alpha\beta})^2}.
\end{equation}
The strange quark contribution to the magnetic moment of the hyperons at tree level, expressed in
terms of $\mu_s$, is used in the one loop calculation.

The contribution to the form factor of Fig.~1d is expressed as
\begin{equation}
G_M^{s(1d)}=-\frac{5{\cal C}^2\mu_s}{288\pi^3f_\pi^2}I_{2K}^{N\Sigma^\ast}.
\end{equation}
Fig.~1e and Fig.~1f provide the following contribution to the nucleon strange magnetic form
factor:
\begin{equation}
G_M^{s(1e+1f)}=-\frac{(D-F){\cal
C}\mu_s}{12\pi^3f_\pi^2}I_{5K}^{N\Sigma\Sigma^\ast},
\end{equation}
where
\begin{equation}
I_{5K}^{\alpha\beta\gamma}=\int d\overrightarrow{k}\frac{k^2 u(\overrightarrow{k})^2}
{\omega_K(\overrightarrow{k})(\omega_K(\overrightarrow{k})+\delta^{\alpha\beta})
(\omega_K(\overrightarrow{k})+\delta^{\alpha\gamma}))}.
\end{equation}

In chiral effective theory, there exists a tadpole diagram (diagram 1g) for the strange form factor.
Its contribution to the strange form factor is
expressed as
\begin{equation}
G_M^{s(1g)}=\frac{\mu_D}{32\pi^3f_\pi^2}I_{4K},
\end{equation}
where
\begin{equation}
I_{4K}=\int d\overrightarrow{k}\frac{u(\overrightarrow{k})^2}
{\omega_K(\overrightarrow{k})}.
\end{equation}

The numerical contribution from Fig.~1g is $0.095$~$\mu_D$ which leads to a positive value.
However, in the previous study, we found that the tadpole contribution should not be included if we
define the quenched low energy constant in the case of FRR to be the 3-quark core contribution~\cite{Wang3}.
As explained in Ref.~\cite{Wang3},
the tadpole contribution from the contact term corresponds to the contribution
of diagram (c) in Fig.~1 summed over an infinite set of highly excited
baryon states and phenomenologically this appears to be appropriately
incorporated through Eq.~(44) in Ref.~\cite{Wang3}.
This equation shows that if we include the tadpole contribution, we have to
determine the unknown low energy constant $a_0'$.
Therefore, the total strange magnetic form factor of the nucleon can be written as
\begin{equation}\label{mup}
G_M^s(Q^2)=\sum_{i=a}^f G_M^{s(1i)}(Q^2),
\end{equation}
where with the regulator parameter $\Lambda=0.8$ GeV, the 3-quark core contribution to the strange form factor is zero.

\begin{table}
\caption{Contributions to the strange magnetic moment of the nucleon $G_M^s$ in unit of $\mu_N$ and the total strange magnetic moment.}
\begin{ruledtabular}
\begin{tabular}{ccccccc}
$\Lambda$ (GeV) & 1a & 1b & 1c & 1d & 1e+1f & $G_M^s$   \\ \hline
0.6 & $-0.021$ &  0.004 & $-0.008$ & $0.005$ & $-0.003$ & $-0.024$   \\
0.7 & $-0.034$ &  0.006 & $-0.014$ & $0.008$ & $-0.005$ & $-0.039$   \\
0.8 & $-0.050$ &  0.009 & $-0.021$ & $0.013$ & $-0.009$ & $-0.058$   \\
0.9 & $-0.070$ &  0.013 & $-0.031$ & $0.019$ & $-0.013$ & $-0.082$  \\
1.0 & $-0.094$ &  0.017 & $-0.043$ & $0.027$ & $-0.018$ & $-0.111$  \\
\end{tabular}
\end{ruledtabular}
\end{table}

\section{Numerical results}

In the numerical calculations, the parameters are chosen as $D=0.76$
and $F=0.50$ ($g_A=D+F=1.26$). The coupling constant ${\cal C}$ is
chosen to be $-1.2$ which is the same as used in Ref.~\cite{Jenkins2}.  The
regulator form factor, $u(k)$, could be chosen to be a monopole, dipole or Gaussian
function, any of which would give similar results~\cite{Young2}.  In our
calculations, a dipole form is used:
\begin{equation}
u(k)=\frac1{(1+k^2/\Lambda^2)^2},
\end{equation}
with $\Lambda = 0.8\pm 0.2$ GeV. This choice has been widely applied in the extrapolation of
lattice data for hadron mass, moments, form factors, radii, first moments of GPDs, etc.~\cite{Leinweber3,Young2,Leinweber:2006ug,Wang4,Wang5,Allton:2005fb,Armour:2008ke}.
With this cloice is has been shown that reasonable physical results can be obtained from the quenched lattice
data at both leading and next leading order~\cite{Leinweber:2004tc,Leinweber:2006ug,Wang1,Wang2,Wang3}.
$\Lambda$ around 0.8 GeV is the value required to identify a core contribution that is invariant between quenched and full QCD.
This invariance of the core  supports the assumption that the core contains no strangeness.

\begin{center}
\begin{figure}[tbp]
\includegraphics[scale=0.7]{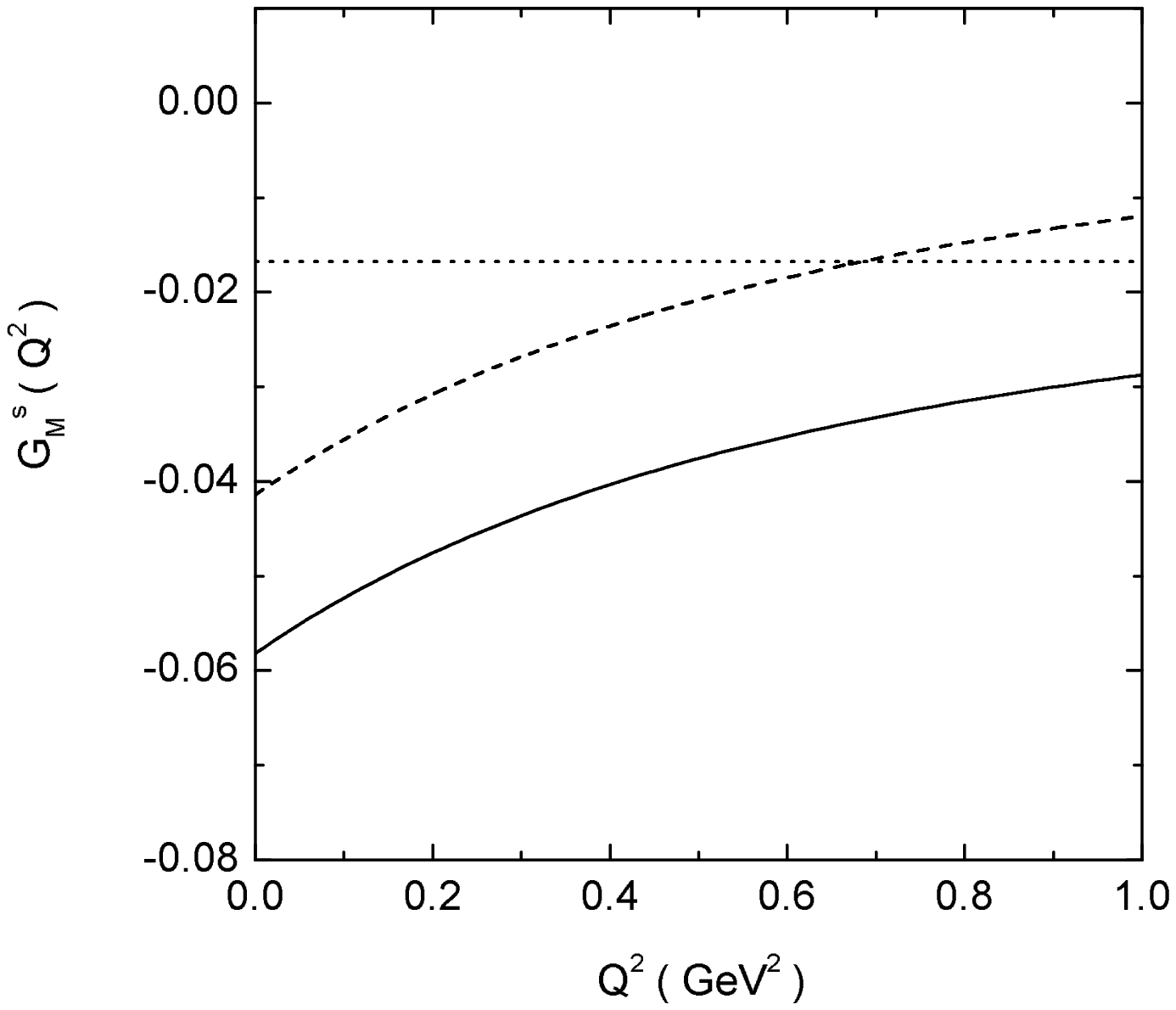}
\caption{The strange magnetic form factor versus $Q^2$. The solid, dashed and dotted lines are for total, leading order and next to leading order results, respectively.}
\end{figure}
\end{center}

\begin{center}
\begin{figure}[tbp]
\includegraphics[scale=0.7]{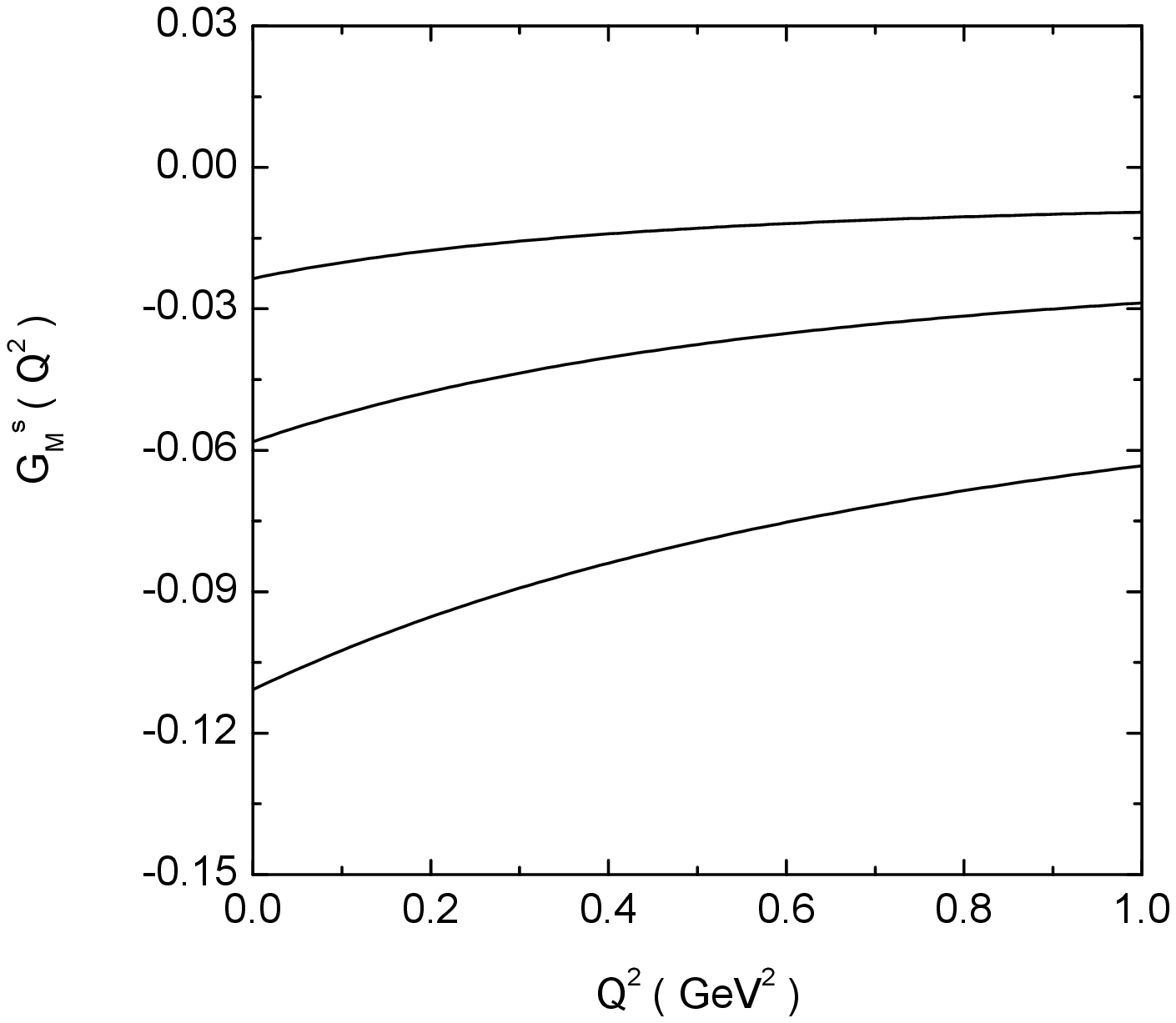}
\caption{The strange magnetic form factor versus $Q^2$. The upper, middle and lower lines are for $\Lambda=0.6$, 0.8 and 1.0 GeV, respectively.}
\end{figure}
\end{center}

The contribution from each diagram in Fig.~1 is shown in Table I.
The value of loop contribution is small, mainly because the $K$ meson mass is large compared with the pion mass.
The leading order diagram (a) gives a negative contribution to the strange magnetic form factor.
Diagram (b), with intermediate decuplet hyperons, gives a positive contribution, but it is numerically $5-6$ times smaller than
that of diagram (a). The sum of the contribution from these two diagrams is $-0.041$~$\mu_N$ for $\Lambda=0.8$ GeV.

The contribution for the next to leading order diagrams, (c), (d), (e) and (f) are much smaller than the leading
contribution. They depend on the parameter $\mu_s$,
which is the constituent strange quark contribution to the hyperon magnetic form factor. Assuming SU(3) symmetry, one has
$\mu_s=\mu_d=-\frac{1}{2}\mu_u=-\frac{1}{3}\mu_D$. In fact, this relation was applied in our previous investigation of nucleon
magnetic form factors \cite{Wang3,Wang4}. In the previous extrapolation of nucleon magnetic form factors, we found
$\mu_D$ equal $2.55$~$\mu_N$ and $2.34$~$\mu_N$ for full QCD and quenched extrapolation, respectively~\cite{Wang3,Wang4}.
Therefore, $\mu_s=-0.8$~$\mu_N$ should be a good estimate. Similar to the leading order case, at next to leading order, the contributions from
intermediate octet and decuplet hyperons (diagrams (c) and (d)) have the opposite sign. For $\Lambda=0.8$ GeV,
the total next to leading order contribution, including transition diagrams (e) and (f), is $0.021$ $\mu_s$ $= -0.017$ $\mu_N$.

At $Q^2=0$, the nucleon strange magnetic moment is $G_M^s(0)=-0.041~\mu_N+0.021~\mu_s$. With the estimate $\mu_s=-0.8$ $\mu_N$, the strange magnetic moment is
$G_M^s(0)=-0.058$~$\mu_N$. If we vary $\Lambda$ from 0.6 GeV to 1 GeV, $G_M^s(0)$ will change from $-0.024$~$\mu_N$ to $-0.111$~$\mu_N$. Numerical results
show that $G_M^s(0)$ is always negative with a large parameter range.

At $\Lambda=0.8$ GeV, the strange magnetic form factor is shown in Fig.~2. The solid, dashed and dotted lines are for total, leading order
and next to leading order results, respectively. The $Q^2$ dependence of $G_M^s(Q^2)$ is determined by the leading order contribution.
The next to leading order contribution is small and $Q^2$ independent, showing that FRR provides good convergence.
Considering the $SU(3)$ symmetry breaking, the next to leading order contribution is even smaller with the smaller $\mu_s$.

In Fig.~3, we plot the strange magnetic form factor $G_M^s(Q^2)$ versus $Q^2$ at $\Lambda=$ 0.6, 0.8 and 1.0 GeV.
One can see that $G_M^s(Q^2)$ decreases in magnitude with the increasing $Q^2$. It is obvious that the strange magnetic form factor does not change sign
for any of the choices of $\Lambda$ when $Q^2$ increases.

With the FRR, the strange form factor is totally determined by the loop contribution. The low energy constants for the 3-quark, tree level contribution
is zero. Therefore, we can calculate the strange magnetic form factor at relatively large $Q^2$. At $Q^2=0$, the strange magnetic moment
is $-0.058^{+0.034}_{-0.053}$~$\mu_N$, in excellent agreement with the lattice QCD results of Refs.~\cite{Leinweber:2004tc,Leinweber:2006ug,Doi:2009sq}.
The numerical values of strange magnetic moment and form factor at different $Q^2$ which the experiments
are focusing on are listed in Table II. The error bars are caused by the uncertainty in $\Lambda$.

\begin{table}
\caption{The strange magnetic form factor at different $Q^2$. Uncertainties reflect the range of $\Lambda$ considered herein.}
\begin{ruledtabular}
\begin{tabular}{cccccc}
$Q^2$ (GeV$^2$)  & 0 & 0.1 & 0.23 & 0.477  & 0.62  \\ \hline
$G_M^s (Q^2)$ & $-0.058^{+0.034}_{-0.053}$ & $-0.052^{+0.031}_{-0.051}$ & $-0.046^{+0.029}_{-0.048}$ & $-0.038^{+0.024}_{-0.040}$ & $-0.035^{+0.023}_{-0.040}$  \\
\end{tabular}
\end{ruledtabular}
\end{table}

\section{Summary}

We have calculated the nucleon strange magnetic form factor within the heavy baryon chiral effective approach using the finite-range-regularization
method up to next to leading order. With the regulator parameter $\sim 0.8$ GeV, the invariance of the core contribution between quenched
and full QCD indicates the low energy constant related to the 3-quark core
contribution is approximately zero. This makes it possible to predict
the strange magnetic form factor, as there is no low energy constant to be specified. The
strange magnetic form factor is totally determined by the FRR loop contribution. The coupling constants in the loop contribution
are well known. The uncertainty arises mostly from the variation of $\Lambda$, which is allowed to run over
the wide range $0.6-1.0$ GeV, as suggested by numerous studies of lattice QCD data as a function of quark mass.

In contrast with the case of the nucleon magnetic form factor, here we do not need to deal with the $Q^2$ dependence of a tree level contribution
to the strange magnetic form factor. As a result, the strange magnetic form factor can be calculated at relatively
large $Q^2$. Numerical results show that the nucleon strange magnetic form factor is always negative, regardless of the variation in the regulator mass.
It decreases in magnitude with increasing $Q^2$, which makes it more difficult for the experiments to measure a nontrivial value for the strange
magnetic form factor at large $Q^2$.

We provide a relatively precise prediction for the nucleon strange form factor.
Our result is consistent with current lattice simulations. The small negative value for the strange magnetic
form factor is also comparable with the existing experimental data.
We look forward to a new generation of precise PVES experiments to compare with our calculated results.

\section*{Acknowledgments}

This work is supported in part by DFG and NSFC (CRC 110), by the
National Natural Science Foundation of China (Grant No. 11035006) and
by the Australian Research Council through grants FL0992247 (AWT),
DP110101265 (DBL) and through the ARC Centre of Excellence for
Particle Physics at the Terascale.

\end{document}